\documentclass[twocolumn]{webofc}

\newcommand{\diff}[1]{\operatorname{d}\ifthenelse{\equal{#1}{}}{\,}{\!#1}}

\newcommand{\MDM}{{\mathchoice{}{}{\scriptscriptstyle}{}\text{MDM}}}
\newcommand{\EDM}{{\mathchoice{}{}{\scriptscriptstyle}{}\text{EDM}}}
\newcommand{\WF}{{\mathchoice{}{}{\scriptscriptstyle}{}\text{WF}}}

\usepackage[varg]{txfonts}   
\usepackage[breaklinks=true,hidelinks]{hyperref}
\usepackage{textcomp}   
\usepackage{amsmath}
\usepackage{graphicx,epsfig}%
\usepackage[shortlabels]{enumitem} 
\usepackage{booktabs} 
\usepackage{bigints} 
\usepackage{relsize}
\usepackage{bm}

\usepackage{breqn}

\begin{document}
\title{Systematic effects in the search for the muon electric dipole moment using the frozen-spin technique}
%
%

\author{%
	\firstname{Chavdar} \lastname{Dutsov}\inst{1}\fnsep\thanks{\email{chavdar.dutsov@psi.ch}} \and
	\firstname{Timothy} \lastname{Hume}\inst{1} \and
	\firstname{Philipp} \lastname{Schmidt-Wellenburg}\inst{1}	
	~on behalf of the muEDM collaboration
}

\institute{Paul Scherrer Institute, Villigen 5232, Switzerland}

\abstract{%
	At the Paul Scherrer Institute (PSI) we are developing a high precision
	instrument to measure the muon electric dipole moment (EDM). The experiment
	is based on the frozen-spin method in which the spin precession induced by
	the anomalous magnetic moment is suppressed, thus increasing the
	signal-to-noise ratio for EDM signals to achieve a sensitivity otherwise
	unattainable using conventional $g-2$ muon storage rings. The expected statistical sensitivity for the EDM after a year of
	data taking is $6 \times 10^{-23} e\cdot$cm with the $p = 125$~MeV/c muon
	beam available at the PSI. Reaching this goal necessitates a
	comprehensive analysis on spurious effects that mimic the EDM signal. This
	work discusses a quantitative approach to study systematic effects for the
	frozen-spin method when searching for the muon EDM. Equations for the motion
	of the muon spin in the electromagnetic fields of the experimental system
	are analytically derived and validated by simulation.}%
\maketitle
\section{Introduction}
The presence of a permanent EDM in an elementary particle implies Charge-Parity
(CP) symmetry violation. Even though the phase of the CKM matrix of the Standard
Model of particle physics (SM) provides a large CP violating phase it results in
tiny electric dipole moments of elementary particles, too small to measure any
time soon.
However, many SM extensions permit large CP violating phases, which also
result in large electric dipole moments~\cite{Pospelov2005, Seng2015PRC}.
Recently, the muon electric dipole moment has become a topic of particular
interest due to the tensions between the measured muon anomalous
magnetic moment and the SM expectation~\cite{Abi2021} and hints of lepton-flavor
universality violation in B-meson decays~\cite{Altmannshofer2017PRD,
	Capdevila2018JHEP}.

The frozen-spin technique, proposed by Farley et al.~\cite{Farley2004}, is a
method of measuring EDMs in storage rings where a radial electric field is
applied to the stored particles such that the anomalous $(g-2)$ precession is
cancelled, and any residual precession is due to the EDM. However, in a real
storage ring the anomalous precession cannot be perfectly negated and EDM-like
precession can be induced by coupling of the magnetic dipole moment (MDM) to the
electromagnetic (EM) fields of the experimental setup -- an example of a
systematic effect.

We define systematic effects as all phenomena that lead to a real or
apparent precession of the spin that are not related to the EDM.

\section{Experimental setup}
The search for the muon EDM will rely on a centimeter-scale storage ring
inside a compact solenoid~\cite{Adelmann2021arXiv}. The muons will be injected
into the solenoid one-by-one through a superconducting injection
channel~\cite{Barna2017} and will be kicked by a pulsed magnetic field into a stable
orbit within a weakly focusing field~\cite{Iinuma2016NIMA}. The muon orbit will
be positioned between two concentric cylindrical electrodes that will provide
the radial electric field to deploy the frozen-spin technique. The direction of
the muon spin at the time of its decay will be deduced from the direction of the
emitted decay positron. This is made possible by their preferential emission,
for high positron energies, in the direction of the muon spin due to parity
violation in the weak decay. Detectors positioned around the ideal orbit will
monitor the direction of emitted positrons. The EDM will be deduced from the
observed change in asymmetry, $A = (N_u - N_d)/(N_u + N_d)$, between upstream
and downstream detectors with time.

The collaboration is proceeding in a staged approach. In the
initial phase we will demonstrate the feasibility of all critical techniques and
aim for sensitivity better than $3\times10^{-21}~e\cdot$cm. In the final phase
we target a sensitivity of better than $6\times 10^{-23} e\cdot$cm -- an
improvement of more than three orders of magnitude over the
current experimental limit $1.8 \times 10^{-19} e\cdot$cm (CL
95\%)~\cite{Bennett2009PRD}.

\section{Limits on systematic effects}
In order to evaluate systematic effects related to the EM fields in the
experiment it is necessary to study the relativistic spin motion in electric
$\vec{E}$ and magnetic $\vec{B}$ fields described by the
Thomas-Bargmann-Michel-Telegdi (T-BMT) equation~\cite{Thomas1927,Montague1984}:
\begin{multline}
	\vec\Omega = -\frac{e}{m_0}\left[ a\vec B - a\frac{\gamma-1}{\gamma}\frac{\left(
			\vec\beta\cdot \vec B \right) \vec \beta
		}{\lvert\vec\beta\rvert^2}\right. +\\
		\left.\left(\frac{1}{\gamma^2-1}-a\right)\frac{\vec \beta \times \vec
			E}{c} + \frac{\eta}{2}\left( \frac{\vec E}{c}+\vec\beta \times \vec B
		\right) \right].
	\label{eq:tbmt_full}
\end{multline}

\begin{figure}[h]
	\centering
	\includegraphics[width=\columnwidth, trim={1.5cm, 2cm, 6cm,
				6cm},clip]{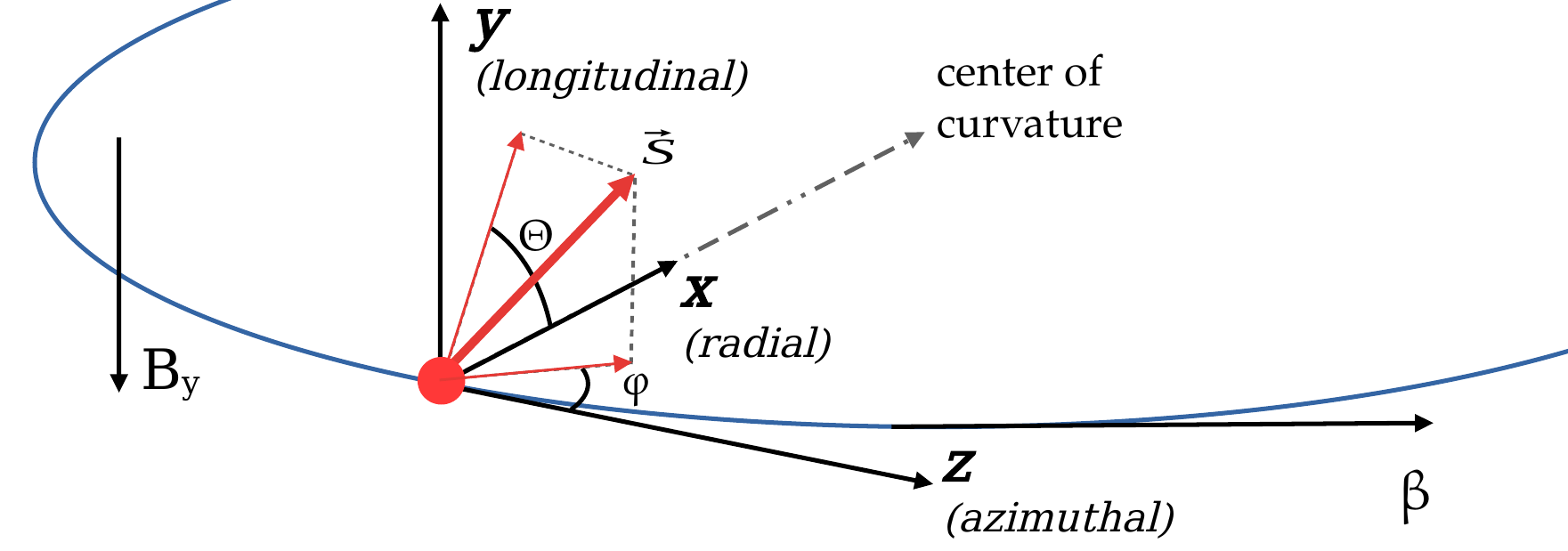}
	\caption{The local reference coordinate system used to derive the motion of
		the spin in the EM fields of the experiment. The vector $\hat z$ follows the
		momentum of the muon and $\hat y$ always points along the longitudinal direction
		defined by the main solenoid magnetic field $B_y$.}
	\label{fig:coord_system}
\end{figure}

The coordinate system used throughout this work is such that it follows the
reference particle orbit (similar to~\cite{HajTahar2021, Rabi1954}) and is
sketched in \autoref{fig:coord_system}.

Using equation \eqref{eq:tbmt_full} and assuming that $\vec \beta \cdot \vec E = 0$ and
$\vec \beta \cdot \vec B = 0$ the angle of spin precession around the radial
axis due to a non-zero EDM is:
\begin{equation}
	\Theta(t) = \Omega^\EDM_{\hat x} t = \frac{\eta}{2}\frac{e}{m_0}\beta_z B_y t,
	\label{eq:edm_angular_velocity}
\end{equation}
where $t$ is the elapsed time between the injection of the muon into the longitudinal
magnetic field and its decay.



The relationship between the dimensionless parameter $\eta$ that characterizes
the spin precession and the EDM $d$ is given by:
\begin{equation}
	d = \frac{e\hslash}{4m_0c}\eta.
	\label{eq:eta_edm}
\end{equation}
Combining \eqref{eq:edm_angular_velocity} and \eqref{eq:eta_edm} gives 
an expression of the rate of change of the angle of rotation of the spin
$\dot \Theta$ as a function of the EDM $d$:
\begin{equation}
	\dot \Theta =  \frac{2c}{\hslash}\beta_z B_y d.
	\label{eq:edm_angle_vs_de}
\end{equation}
For the precursor and final experiments $\beta_z = 0.26$ and $0.77$ respectively
and $B_y = 3$ T, hence, the angular velocities are
\begin{align}
	\dot \Theta & = 21.15 \text{ \textmu rad/\textmu s}, \text{ for } d =
	3\times 10^{-21}~e\cdot\text{cm and}
	\label{eq:limit_on_angular_velocity_prec}                              \\
	\dot \Theta & = 1.26 \text{ \textmu rad/\textmu s}, \text{ for } d =
	6\times 10^{-23}~e\cdot\text{cm}.
	\label{eq:limit_on_angular_velocity}
\end{align}

In the following sections we will discuss various effects with
the goal of deriving the full description of the spin's motion in the
EM fields of the system. The derived equations can be used to place limits on
measurement parameters such that the angular velocity of the spin precession due
to the MDM is less than the experimental sensitivities shown in
\eqref{eq:limit_on_angular_velocity_prec} and \eqref{eq:limit_on_angular_velocity}.

\section{Sources of real spin precession}
One of the large sources of a false EDM signal are oscillations and rotations of
the spin due to the coupling of the MDM of the muon with the electric and
magnetic fields present in the experimental setup. As a starting point we will
describe analytically the motion of the spin due to the combined effect of ideal
approximations of the magnetic field of the solenoid, the weakly focusing field
and the electric field used in the frozen-spin method. The magnetic field of the
solenoid is approximated as a uniform magnetic field oriented along the
$\hat y$ axis. The weakly focusing field is described by the approximated field
that is generated by a circular coil. The electric field is assumed to be a
radial field generated from the potential difference between two infinite
coaxial cylindrical electrodes. Some possible and important imperfections of
these fields and their effect on the spin precession is discussed.

\subsection{Spin precession along the radial axis}\label{sec:weakly_focusing_desc}
Assuming the particles follow a trajectory with a constant radius $\rho$,
the radial magnetic field $B_x(y)$ of a cylindrical coil with $N$ turns and radius
$R$ along that trajectory can be approximated with~\cite{Murgatroyd1991}:
\begin{equation}
	B_x(y) = \frac{3}{2}\mu_0NIR^2 \frac{\rho y}{\left( R^2 + y^2
	\right)^{\frac{5}{2}}},
	\label{eq:radial_b_weakly_focus}
\end{equation}
where $I$ is the current passing through the coil and $\mu_0$ is the
magnetic permittivity of vacuum. Expanding \eqref{eq:radial_b_weakly_focus} in
a Taylor series one obtains:
\begin{equation}
	B_x(y) = \Phi_0 \rho y - \frac{5}{2} \frac{\Phi_0}{R^2} \rho y^3 +
	\mathcal O(y^5), 
	\text{ where }
	\Phi_{0} = \frac{3}{2}\frac{\mu_0 NI}{R^3}.
	\label{eq:radial_b_weakly_focus_taylor}
\end{equation}

The longitudinal position of a particle with charge $e$, mass $m_0$ and velocity
$c\vec\beta$ is given by the solution of:
\begin{equation}
	\ddot{y} = \frac{1}{\gamma m_0} \left(e E_y + c \beta_z B_x(y)\right),
	\label{eq:lorentz}
\end{equation}
where we assume a constant non-zero longitudinal component of
the electric field. The solution of the differential equation is the harmonic
oscillator:
\begin{align}
	 & y(t) = y_0\cos (\omega_\beta t + \varphi) +
	\frac{1}{\Phi_0\rho}\frac{E_y}{c\beta_z},       \\ &\text{where } \omega_\beta = \sqrt{\Phi_{0}\frac{ec\beta_z}{\gamma m_0} \rho}
	\label{eq:omega_betatron}
\end{align}
is the angular velocity of the vertical betatron oscillations (VBO).

For the precursor experiment where $\beta = 0.258$ and $\gamma = 1.03$ the period $T
	= 2\pi/\omega_\beta$ of the VBO is approximately 600~ns. Note 
that this period depends only on the strength of the focusing field and the
radial position of the muon. It does not depend on the momentum of the muon.

The precession of the spin due to the coupling of the MDM with the radial
magnetic field due to the weakly focusing field is: 
\begin{multline}
	\left(\vec\Omega_\MDM^\WF \right)_x = -\frac{ea}{m_0}B_x(y(t)) = \\
	= -\frac{ea}{m_0}\left[\Phi_{0}\cos(\omega_\beta t + \phi_0)\rho y_0 - \frac{1}{c\beta_z}
		E_y\right].
	\label{eq:precession_weakly_focusing}
\end{multline}

The other source of radial precession that has to be considered is the
radial magnetic field in the reference frame of the muon due to the non-zero
longitudinal electric field in the laboratory reference frame. Using the T-BMT
equation and taking only the radial component of the spin precession due to the
electric field one obtains:
\begin{equation}
	\left(\vec\Omega_\MDM^{\scriptscriptstyle{}E_y}\right)_x
	=
	-\frac{e}{m_0c} \left( a - \frac{1}{\gamma^2 -1} \right)
	\beta_z E_y.
	\label{eq:precession_ey}
\end{equation}
Combining \eqref{eq:precession_weakly_focusing} and
\eqref{eq:precession_ey}, for the total angular velocity of the
radial precession due to the MDM around the $\hat x$ axis, one obtains :
\begin{multline}
	\left( \vec\Omega_\MDM \right)_x =
	-\frac{ea}{m_0} \left[
		\frac{1}{c}\left( 1-\frac{1}{a(\gamma^2 - 1)} -
		\frac{1}{\beta^2_z}\right) \beta_z E_y\right. + \\
		\left.\Phi_{0}\cos(\omega_\beta t + \phi_0) \rho y_0 \right].
	\label{eq:precession_mdm_total}
\end{multline}

\subsection{Azimuthal spin precession}
When the muons circulate in the storage ring they oscillate around an
equilibrium orbit. Because of this oscillation the momentum of the particle is
not at all times perpendicular to the longitudinal magnetic field leading to a
non-zero projection of the magnetic field along its trajectory. This field is
proportional to the angle between the muon momentum and the magnetic field
$\delta = \angle (\vec\beta, \vec B)$. In turn, $\cos(\delta) = p_y / p_z$ and
$p_y$ oscillates as $p_y = p_{y_0} \sin(\omega_\beta t)$. Thus:
\begin{equation}
	B_z(t) = B_y \cos(\delta) \approx \frac{p_{y_0}}{p_{z}} \sin(\omega_\beta t) B_y.
	\label{eq:b_theta}
\end{equation}
The momentum $p_{y_0}$ is the momentum in the $\hat y$ direction when the muon is on
the equilibrium orbit. It can be calculated as:
\begin{equation}
	p_{y_0} = e c \beta \Phi_0 \rho y_0 \int_0^{\frac{\pi}{2 \omega_\beta}}
	\cos(\omega_\beta t ) dt = e c \beta \Phi_0 \rho y_0 \frac{1}{\omega_\beta}.
	\label{eq:p_y_0}
\end{equation}

In the ideal case $\vec \beta = \lvert\beta\rvert \hat z$ and $\vec E = \lvert
	E\rvert\hat x$, but as the muons oscillate in the weakly focusing field there
will be a non-zero $\hat y$ component of the velocity, thus:
\begin{equation}
	\beta_y = (p_{y_0}/p_{z})\,\beta_z \sin(\omega_\beta t).
	\label{eq:beta_y}
\end{equation}

If the radial electric field $E_x$ is correctly set to the value $E_f$ required
by the frozen-spin technique, then there will be no oscillations around the
$\hat z$ axis, excluding the second order $\beta \cdot \vec B$ term in
\eqref{eq:tbmt_full}. At this value for the freeze field, it will counteract the
precession induced by coupling of the MDM to the longitudinal field of the
solenoid. However, if $E_x \neq E_f$ there will be imperfect cancellation of the
$g-2$ precession around $\hat z$ which is proportional to the excess electric
field $E_{ex} = E^{(\text{real})} - E^{(\text{freeze})}$ that the muon is
subjected to:
\begin{equation}
	\left(\vec \Omega_\MDM^{\scriptscriptstyle E_{ex}} \right)_z = -\frac{e}{m_0c}
	\left( a - \frac{1}{\gamma^2 - 1} \right)\beta_y E_{\text{ex}}.
	\label{eq:precession_mdm_excess_efield}
\end{equation}

Considering \eqref{eq:b_theta}, the angular velocity of the spin precession along
the $\hat z$ axis due to the $\vec\beta \cdot \vec B$ term in the T-BMT equation
is:
\begin{equation}
	\left( \vec\Omega_\MDM^{\,\scriptscriptstyle\beta\cdot B} \right)_z = \frac{ea}{m_0} \left(\frac{\gamma -
		1}{\gamma}\right) B_z(t).
	\label{eq:precession_mdm_theta}
\end{equation}

Combining equations \eqref{eq:precession_mdm_theta} and
\eqref{eq:precession_mdm_excess_efield} gives the total angular velocity of the
precession due to the MDM around the $\hat z$ axis:
\begin{multline}
	\left(\vec\Omega_\MDM\right)_{z} =
	-\frac{e}{m_0}\frac{p_{y_0}}{p_z}
	\sin\left( \omega_\beta t \right) \times  \\
	\left[ \left( a - \frac{1}{\gamma^2 - 1} \right)
	\frac{\beta_z}{c}E_{ex} -
	a \left( \frac{\gamma - 1}{\gamma} \right)B_y \right].
	\label{eq:precession_mdm_z_oscillations}
\end{multline}

\subsection{Electric field imperfections}
The value of $E^{(\text{real})}$ may not be constant throughout the orbit of the
muon if the central axes of the magnetic and electric fields are displaced or
inclined with respect to each other. To obtain the components of the electric
field with respect to the central axis of the solenoid, first consider a purely
radial electric field created by perfect coaxial cylindrical electrodes with
radii $A$ and $B$. In the reference frame $(x', y', z')$ of the electrodes,
where $z'$ is parallel to the central axis of the electrodes, the field is given
by:
\begin{equation}
	\vec E'(\vec r) = \frac{V}{\log{\frac{B}{A}}}\begin{pmatrix}x'/r^2 \\y'/r^2
		\\0\\\end{pmatrix},
	\label{eq:e_field_vector}
\end{equation}
where $r^2 = (x')^2 + (y')^2$. The electrode central axis, however, may be
displaced and rotated with respect to the reference frame defined by the
solenoid central axis. The electric field $\vec E$ in the reference frame of the solenoid
can be obtained by transforming $\vec E'$ as:
\begin{equation}
	\vec E = R_y(\alpha) \vec E'(R_y^{-1}(\alpha)\vec r + \vec r_0),
	\label{eq:e_field_ref_frame}
\end{equation}
where $\vec r_0 = (x'_0, y'_0, 0)$ is the displacement between the two fields.
$R_y(\alpha)$ is the rotation around the $y'$ axis at an angle~$\alpha$, where
$\alpha$ is the angle between the central axis of the cylindrical electrodes and
the central axis of the longitudinal magnetic field.
Note here that due to the rotational symmetry of the fields we can always
choose the reference frame to be such that arbitrary displacements can be
represented in this way. The electric field in the reference frame
defined by the longitudinal magnetic field is then:
\begin{equation}
	\vec E(\vec r) = V_0
	\begin{pmatrix}
		\frac{\xi}{\rho^2} \cos\alpha \\
		\frac{\upsilon}{\rho^2}       \\
		-\frac{\xi}{\rho^2} \sin \alpha
	\end{pmatrix}
	\label{eq:e_field_rotated}
\end{equation}
where $V_0 = V/\log{(B/A)}$, $\upsilon = y' + y'_0$, $\xi = x'_0 +
	x' \cos\alpha - z' \sin\alpha$ and $\rho^2 = \xi^2 + \upsilon^2$.

The average of the radial electric field over the circular orbit of the muon can
be obtained if $E'$ is represented in cylindrical coordinates $(\rho, \phi,
	\zeta)$ as:
\begin{equation}
	\left\langle E(\rho, \zeta) \right\rangle = \frac{1}{2 \pi} \int_0^{2\pi} E d\phi,
	\label{eq:average_e_field}
\end{equation}
where $\rho$ is the radius of the orbit of the muon as a function of the
magnetic field and the momentum of the particle, and $\zeta$ is parallel to
$\hat y$.

The radial electric field the muon experiences in its reference frame is modeled as:
\begin{equation}
	E^{(\text{real})} = \left\langle E(\rho,\zeta) \right\rangle_\rho +
	\frac{1}{2}\left(E_{\rho, max} - E_{\rho, min}\right)\cos (\omega_c t +
	\beta_0),
	\label{eq:model_e_field_cyclotron}
\end{equation}
where $\omega_c = -eB_y/\gamma m_0$ is the cyclotron angular velocity, $E_{max}$ and
$E_{min}$ are the maximal and minimal value of the electric field over one
rotation of the muon and $\beta_0$ is the initial phase of the muon position
along the orbit.

Note, \eqref{eq:average_e_field} is valid only in the case of a
circular orbit. In this case it can be shown numerically that:
\begin{equation}
	\left\langle E(\rho, \zeta) \right\rangle = \left\langle E'(\rho,
	\zeta)\right\rangle,
	\label{eq:rotation_orbit_doesnt_matter}
\end{equation}
i.e., the rotation of the anode or cathode with respect to the muon orbit does
not influence the average frozen spin condition and, more importantly, does not
change the net E-field component in the $\hat y$ direction. As the centripetal
force due to the B-field is $\approx10^3$ higher than that due to the E-field
and the expected misalignments between the center of the orbit and the center of
the inner electrode are small, the circular orbit approximation holds well.

\subsection{Combined spin precession}
The initial orientation of the spin $\vec S = (S_x, S_y, S_z)$ in spherical
coordinates is: 
\begin{equation}
	\phi_0   = \arctan\left( \frac{S_x}{S_z} \right), \quad
	\Theta_0 = \arctan\left( \frac{\sqrt{S^2_x + S^2_z}}{S_y} \right) -
	\frac{\pi}{2}.
	\label{eq:phi_theta_initial}
\end{equation}
If there is an imperfect cancellation of the $g-2$ precession then there will be a
rotation of the spin around the $\hat y$ axis. Thus, the angular velocity of
the spin precession around $\hat x$ and $\hat z$ will be a projection of
\eqref{eq:precession_mdm_z_oscillations} and \eqref{eq:precession_z} along
$\hat y$:
\begin{equation}
	\lvert\vec\Omega_\MDM\cdot\hat y \rvert = \left( \vec\Omega_\MDM \right)_x\cos(\omega_y t + \phi_0) +  \left(\vec\Omega_\MDM
	\right)_z \sin(\omega_y t + \phi_0),
	\label{eq:projection_omega_mdm}
\end{equation}
where $\omega_y$ is the angular velocity of the precession around $\hat y$ due
to the $g-2$ precession and is:
\begin{equation}
	\omega_y = \frac{ea}{m_0}\frac{E_{\text{ex}}}{E_f} B_y.
	\label{eq:omega_g-2}
\end{equation}
In practice, $E_{\text{ex}}$ will oscillate (see
\eqref{eq:model_e_field_cyclotron}) with the cyclotron frequency due to the changing
distance between the muon and the E-field center. The longitudinal B-field $B_y$
will oscillate with the VBO frequency due to the changing longitudinal
component of the weakly focusing field (proportional to
\eqref{eq:radial_b_weakly_focus}). In a well tuned frozen-spin experiment
$\omega_y$ is much lower than $\omega_\beta$ and $\omega_c$, and in this
derivation the rotation of the spin around $\hat y$ is approximated with
a constant angular velocity using the average values of $E_{\text{ex}}$ and
$B_y$. 

The total longitudinal rotation of the spin $\Theta$ is:
\begin{equation}
	\Theta(t) = \int_0^t \lvert\Omega_\MDM\cdot\hat y \rvert dt
	\label{eq:precession_theta_integral}
\end{equation}
Calculating \eqref{eq:precession_theta_integral} for the $\hat z$ component
only one obtains:
\begin{equation}
	\begin{split}
		\Theta_z (t) &= -\frac{e}{m_0} \mathop{\mathlarger{\int}}_0^t\left[ \frac{p_{y_0}}{p_z} \left( \frac{\beta_z}{c} \left( a - \frac{1}{\gamma^2 - 1} \right)E_{\text{ex}}\right.\right.\\
			& - \left. \left.a \left( \frac{\gamma - 1}{\gamma} \right)B_y \right)\sin(\omega_\beta t') + aB_z\right] \sin(\omega_y t' + \phi_0) dt'
		\label{eq:theta_x_inegral_calculation_step_1}
	\end{split}
\end{equation}
Noting that:
\begin{multline}
	\int_0^t \sin(\omega_\beta t')\sin(\omega_y t' + \phi_0) dt' \approx \\
	-\frac{1}{\omega_\beta} \cos(\omega_\beta t)\sin(\omega_y t +
	\phi_0),\quad\text{if }\omega_\beta >> \omega_y,
	\label{eq:approx_int_sin_sin}
\end{multline}
the oscillations due to the excess electric field along the $\hat z$ axis (along the
momentum) and due to the non-zero $B_z$ magnetic field can be calculated by:
\begin{equation}
	\begin{split}
		\Theta_{\hat z}(t) &=
		\frac{e}{m_0}\left[\frac{p_{y_0}}{p_z}\frac{1}{\omega_\beta}
			\cos\left( \omega_\beta t \right) \left(
			\frac{\beta_z}{c}\left( a - \frac{1}{\gamma^2 - 1} \right)
			E_{\text{ex}}\right.\right. \\ 
			&- \left.\left.a \left( \frac{\gamma - 1}{\gamma} \right)B_y\right) - \frac{1}{\omega_y} aB_z \right]\sin(\omega_y t + \phi_0).
		\label{eq:precession_z}
	\end{split}
\end{equation}

A similar calculation can be
performed for the spin-precession along the $\hat x$ axis due to the radial
magnetic field of the weakly focusing field:
\begin{equation}
	\Theta_{\hat x}(t) =
	-\frac{ea}{m_0} \left[
		\frac{1}{\omega_\beta}\Phi_{0}\sin(\omega_\beta t + \phi_0) \rho y_0
		\right]\cos(\omega_y t + \phi_0).
	\label{eq:precession_x}
\end{equation}

The spin-precession due to a non-zero electric field in the $\hat y$ direction
can be easily integrated without approximations and gives:
\begin{multline}
	\Theta_{\hat z \times \hat y}(t) =
	-\frac{ea}{m_0} \left[
		\left(1 -  \frac{1}{a(\gamma^2 - 1)} -
		\frac{1}{\beta_z^2}\right)\frac{\beta_z}{c}E_y \right] \times \\
	\frac{1}{\omega_y} \sin(\omega_y t + \phi_0).
	\label{eq:precession_x_v}
\end{multline}

The total azimuthal angle $\Theta$ is the sum of the oscillations and rotations
given above:
\begin{equation}
	\Theta = \Theta_{\hat x} + \Theta_{\hat z} + \Theta_{\hat z \times \hat y} + \Theta_0.
	\label{eq:precession_angle_mdm_total}
\end{equation}

\paragraph{Input parameters}
The analytical equations have a few input parameters which can be roughly
divided into two groups: stochastic initial conditions that vary for each particle in
the storage ring at time $t_0$ of the experiment and constant or slowly changing parameters of the experimental
system. The stochastic conditions in the first group are:
\begin{equation}
	\begin{aligned}
		\vec p_0 & = (p_x, 0, p_z) - \text{Initial momentum of the particle.}   \\
		\vec S_0 & = (S_x, S_y, S_z) - \text{Initial spin of the particle.}     \\
		\vec r_0 & = (r_x, r_y, r_z) - \text{Initial position of the particle.} \\
		t_0      & - \text{Time after injection for which $p_y = 0$.}           \\
	\end{aligned}
	\label{eq:inital_stochastic_parameters}
\end{equation}
Note here, that the $p_y$ component is equal to zero as this is the defining
moment when the particle is considered stored in the storage ring. The
parameters $r_x$ and $r_z$ as well as $p_x$ and $p_z$ define the orbit of the
muons and determine the electric field $E^{(\text{real})}$ to which they are
subjected. The parameter $r_y$ is the starting position of the muon along the
solenoid axis and defines the oscillations due to the radial weakly focusing
magnetic field. The initial spin direction $\vec S_0$ determines the intermixing
between the rotations around the $\hat x$ and $\hat z$ axes. There is also one
constraint imposed on the spin orientation: $\lvert S_0\rvert = 1$.

The parameters of the experimental system are:
\begin{enumerate}
	\item $B_y$ -- Main solenoid magnetic field strength.
	\item $I, R, N$ -- Current through the $N$ turns of the weakly focusing
	      coils with radius $R$.
	\item $V, b, a, \alpha$ -- Voltage between the concentric cylinders with radii
	      $b$ and $a$ defining the weakly focusing field. The angle $\alpha$
	      between the central axes of the electric field and solenoid
	      magnetic field.
	\item $p^{\scriptscriptstyle\text{(ideal)}}_0$ -- Ideal momentum of the
	      muons for which the freeze field is set.
	\item $E_y$ -- Constant electric field normal to the
	      storage ring plane.
\end{enumerate}
Thus, the analytical equations depend on 8 free stochastic parameters and 10
constant parameters of the experimental system. These parameters completely
define the dynamics of the spin inside this model of a storage ring.

\subsection{Comparison with Geant4 spin tracking}
In order to verify the analytical equations a model of the experiment was
developed using the Geant4 Monte Carlo simulation toolkit~\cite{Agostinelli2003}.
The EM fields of the experiment can be either calculated analytically or
interpolated from field maps. The field maps are generated by ANSYS Maxwell
finite element modeling (FEM) software. The simulation has three major EM field
components:
\begin{enumerate}
	\item The main solenoid magnetic field -- constant value along $\hat y$ or
	      field map supplied by FEM simulations.
	\item Radial electric field given by
	      \eqref{eq:e_field_rotated} with a
	      possibility to add a constant and uniform component in the longitudinal
	      direction or FEM simulated field map.
	\item Weakly focusing field modeled in ANSYS as a single circular coil with
	      $R = 65$~mm.
\end{enumerate}


The muons start with zero momentum in the $\hat y$ direction as this is the
initial condition for a stored muon. The simulation tracks the spin orientation
in the reference frame of the muon and records it as a function of time. It can
also track the direction with respect to a reference frame defined by the
experimental setup, e.g., the solenoid.

\begin{figure}[h]
	\centering
	\includegraphics[width=\columnwidth]{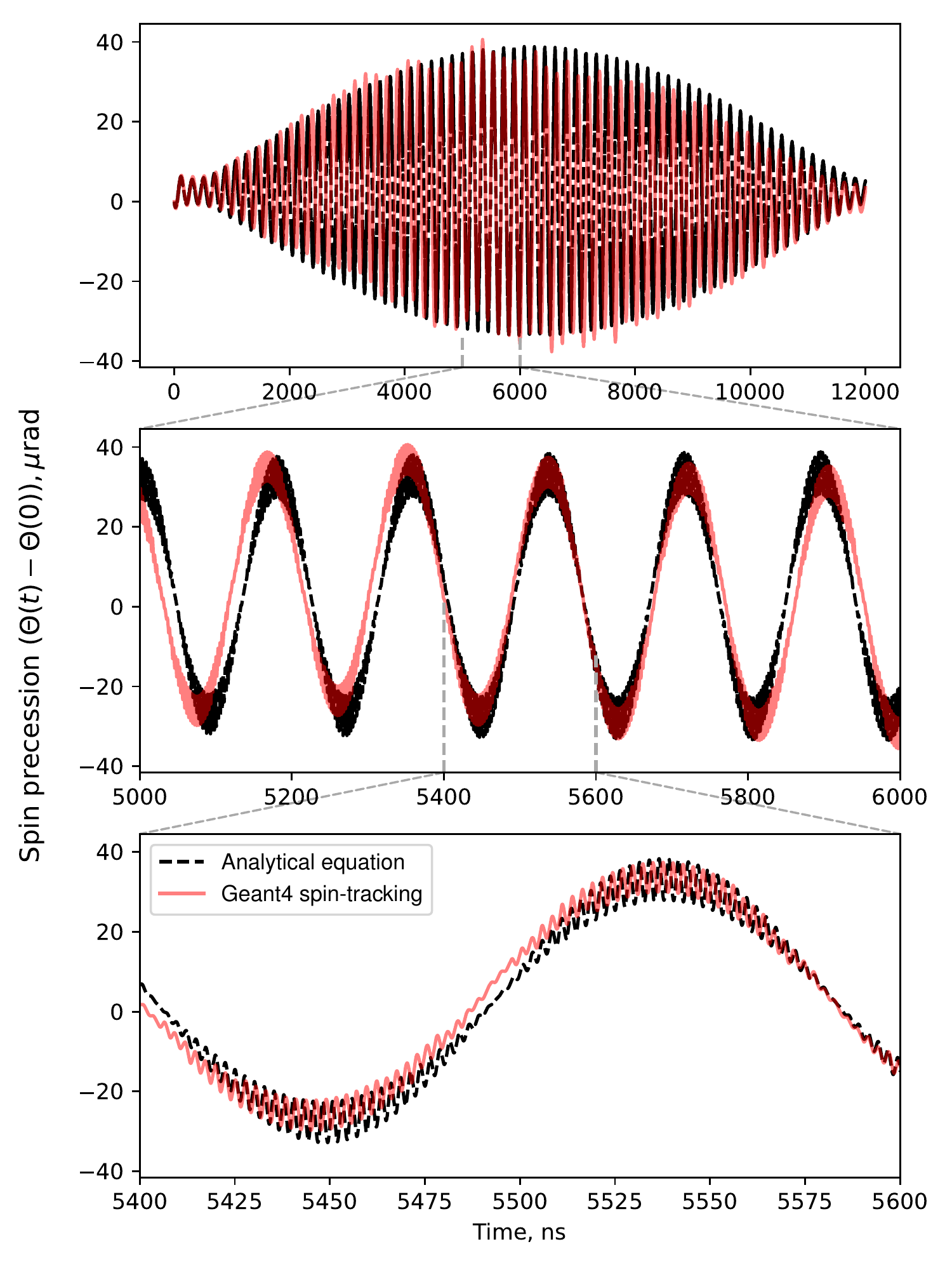}
	\caption{Comparison between the analytical equations
	\eqref{eq:precession_angle_mdm_total} and the Geant4 spin tracking simulation. The
	initial parameters are arbitrarily set to $\vec S_0 = (-0.89, 0.00, -0.46)$,
	$\vec p_0 = (0.84, 0.00, 0.55) \times 26.8$~MeV/c and $\vec r_0 =
		(-16.95, 5.00, 22.10)$~mm. The ideal momentum for which the freeze field
	is set is $p_0^{\scriptscriptstyle\text{(ideal)}} =
		28.0$~MeV/c. The three subfigures show different
	timescales of the spin motion -- uncompensated $(g-2)$ precession,
	betatron oscillations and cyclotron oscillations.}
	\label{fig:good_agreement_ana_g4}
\end{figure}

A comparison between the derived analytical equations and the Geant4 spin
tracking is shown in Figure~\ref{fig:good_agreement_ana_g4}. In the comparison a
fine field map (0.2~mm step size) of the weakly focused field was used. All
other EM fields are calculated analytically. The stepper used is the
DormandPrinceRK78 routine~\cite{Agostinelli2003} with $0.26$~mm step size. The
stepping size was chosen so as to avoid effects due to resonances between the
stepper and field map grids. The initial coordinates of the muon at
the moment of storage were chosen arbitrarily (values specified in
\autoref{fig:good_agreement_ana_g4} caption) for the purpose of the
illustration. The electric field was set to such a value as to have
imperfect cancellation of the $(g-2)$ precession. The anode coaxial E-field is
tilted with respect to the longitudinal axis at $\alpha = 0.01^\circ$ in order
to highlight the cyclotron oscillations.

The comparison shows very good agreement between the analytical equations and
the Geant4 spin tracking. The approximate equation for the description of the
weakly focusing field \eqref{eq:radial_b_weakly_focus_taylor} provides good
estimates of the field strength and VBO frequency. The
difference between the analytical and numerical approaches is less than 5
\textmu rad sustained over 12~\textmu s of simulation time, demonstrating the
very good agreement between the two also when  using realistic FEM generated
field maps in Geant4.




\section{Sources of apparent spin precession}
If the muon spin has a component in the longitudinal direction then the
probability for upstream $p_u$ and downstream $p_d$ ejected decay positrons will
differ and an asymmetry $A$ will be observed:
\begin{equation}
	A(t) = \frac{p_u - p_d}{p_u + p_d} = \sin\left(\Theta(t) \right) \alpha P
	\approx \Theta(t)\, \alpha P,
	\label{eq:asym_theory}
\end{equation}
where $\alpha$ is the parity violating decay asymmetry averaged over the
positron energy and $P$ is the initial polarization. Equation
\eqref{eq:asym_theory} is valid when $\Theta(t) \ll 1$, which is the case of EDM
induced spin precession.

Using \eqref{eq:edm_angular_velocity} and \eqref{eq:asym_theory}, the rate of change
of the asymmetry $\dot A$ is:
\begin{equation}
	\dot A =  \frac{\eta}{2}\frac{e}{m_0} \beta_z B_y \alpha P.
	\label{eq:emission_dir_asymmetry_derivative}
\end{equation}
We would require that the measured asymmetry $\dot A_m$ due to
spurious EDM mimicking effects is $\dot A_m \ll \dot A$.

The number of detected positrons in the upstream detector $N_u$ is given by:
\begin{equation}
	N_u = \Omega_u \varepsilon_u p_u D,
	\label{eq:n_up}
\end{equation}
where $D$ is the total number of decayed muons, $\varepsilon_u$ is the detection
efficiency for positrons and $\Omega_u$ is the solid angle coverage of the
upstream detector. Both $\Omega_u$ and $\varepsilon_u$ can be summarized with a
single parameter $\kappa_u = \Omega_u \varepsilon_u$ expressing the effective
detection efficiency of the upstream detector. A similar equation can be given
for the downstream detector and so:
\begin{equation}
	N_u = \kappa_u p_u D = \kappa_u D_u\text{ and }
	N_d = \kappa_d p_d D.
	\label{eq:n_down}
\end{equation}
From the point of view of the experiment, substituting equation \eqref{eq:n_down} in
\eqref{eq:asym_theory}, the measured asymmetry $A_m$ is given by:
\begin{equation}
	A_m = \frac{1}{D}\left(\frac{N_u}{\kappa_u} -
	\frac{N_d}{\kappa_d}\right).
	\label{eq:measured_asymmetry}
\end{equation}
If the effective detection efficiencies $\kappa$ are constant in time no
systematic effect appears, but a time dependence, e.g., by a
disturbance of the detector electronics through the pulsed kicker field, leads
to a systematic bias. A worst-case scenario assumption is the case where the kicker field
disturbs the detection efficiency in both the upstream and downstream detectors
in opposite directions and its effect reduces exponentially with time. The disturbance can be modeled as:
\begin{equation}
	\kappa_u = \kappa_{u0} - \Delta_\kappa e^{-t/\tau_k}\text{ and } \kappa_d = \kappa_{d0} + \Delta_\kappa e^{-t/\tau_k},
	\label{eq:exponential_kappa}
\end{equation}
where $\kappa_{u0}$ and $\kappa_{d0}$ are some equilibrium detection
efficiencies, $\Delta_\kappa$ is the perturbation in the efficiency due to the
kicker field and $\tau_k$ is the time-constant of its effect.

In order to quantify the false EDM signal that can be caused by time-dependent
efficiency parameters let us assume that there is no true EDM signal and the
upstream and downstream emission probabilities are equal $p_u = p_d$ leading to
$D_u = D_d = D/2$. Then the observed asymmetry is:
\begin{equation}
	A_m = \kappa_u - \kappa_d = \kappa_{u0} - \Delta_\kappa e^{-t/\tau_k} -
	\kappa_{d0} - \Delta_\kappa e^{-t/\tau_k}.
	\label{eq:asymmetry_exponential_kappa_2}
\end{equation}
The measured EDM signal is given by the time derivative of the asymmetry, thus:
\begin{equation}
	\dot A_m = \frac{2}{\tau_k} \Delta_\kappa e^{-t/\tau_k}.
	\label{eq:asymmetry_exponential_kappa_fin}
\end{equation}

From equation \eqref{eq:asymmetry_exponential_kappa_fin} we can observe that the
systematic effect decreases with time $t$ as expected. The effect is exacerbated
if $\tau_k$ is on the order of magnitude of the experimental measurement time
$t$ which is several muon lifetimes. The limits on $\tau_k$ and $\Delta_\kappa$
can be derived by requesting that the measured false asymmetry $\dot A_m$ is
lower than the theoretical prediction $\dot A$ from
\eqref{eq:emission_dir_asymmetry_derivative} for a given $\eta$.

\section{Conclusions}
Analytical equations for the description of the MDM precession in the EM fields
of the experiment were derived. The equations were compared to a Geant4 Monte
Carlo simulation using realistic field maps generated using the FEM software
ANSYS Maxwell. Very good agreement has been observed between the two and the
analytical equations seem to describe well the spin motion at short, medium and
long timescales.

A method of calculation of the measured false EDM signal due to exponential
changes in the positron detection efficiency is shown. If the detection
efficiency is constant, even if different, for each detector for early-to-late
times, then there would be no false EDM signal. A false EDM would be generated
in the case of change in the detection efficiency of upstream relative to
downstream detectors and the systematic effect is more prominent for
time-constants on the order of the muon lifetime.

The presented methods will be used to derive limits on the mechanical production
and EM field uniformity of the muon EDM experiments to be built at the PSI.

\section{Acknowledgements}
This work is financed by the Swiss National Science Fund under grant
\textnumero~204118.

\bibliographystyle{woc.bst}
\bibliography{biblio}
\end{document}